\documentclass[letter,conference]{ieeeconf}%
\IEEEoverridecommandlockouts%
\usepackage{cite}
\usepackage{amsmath,amssymb,amsfonts}
\usepackage{standalone}
\usepackage{float}

\usepackage{varwidth, tikz}
\usetikzlibrary{shapes, arrows, shapes.misc, arrows.meta, positioning, matrix, calc, fit, fadings}
\usepackage{relsize}

\newlength{\lw}
\newlength{\lwt}
\newlength{\lnshift}
\newlength{\sshift}
\newlength{\ssshift}
\newlength{\lhalf}
\newlength{\lone}
\newlength{\lonea}
\newlength{\ltwo}
\newlength{\lthree}
\newlength{\vone}
\newcommand{\st}[1]{\normalsize{#1}}
\newcommand{\ssmp}{\scalebox{0.7}[0.7]{$+$}}
\newcommand{\ssm}{\scalebox{0.7}[0.7]{$-$}}

\newlength{\spaceimcstandard}
\newlength{\imcvspace}
\newlength{\imchspace}

\usepackage{parskip} 
\begingroup
    \makeatletter
    \@for\theoremstyle:=definition,remark,plain\do{%
        \expandafter\g@addto@macro\csname th@\theoremstyle\endcsname{%
            \addtolength\thm@preskip\parskip
            }%
        }
\endgroup

\usepackage[absolute]{textpos}
\usepackage{balance}
\usepackage{graphicx}
\usepackage{overpic}
\usepackage{rotating}
\usepackage{algorithm,algorithmic}
\usepackage{mathtools}

\usepackage{mymath}
\usepackage{nth}

\makeatletter
\renewcommand*\env@matrix[1][\arraystretch]{%
  \edef\arraystretch{#1}%
  \hskip -\arraycolsep
  \let\@ifnextchar\new@ifnextchar
  \array{*\c@MaxMatrixCols c}}
\makeatother

\newtheorem{theorem}{Theorem}

\usepackage{graphicx}
\usepackage{textcomp}
\usepackage{xcolor}
\def\BibTeX{{\rm B\kern-.05em{\sc i\kern-.025em b}\kern-.08em
    T\kern-.1667em\lower.7ex\hbox{E}\kern-.125emX}}

\usepackage{caption}
\usepackage{subcaption}
\begin{document}

\title{\LARGE \bf Multi-Array Electron Beam Stabilization using Block-Circulant Transformation and Generalized Singular Value Decomposition}

\author{Idris Kempf${^*}$, Stephen R. Duncan${^*}$, Paul J.\ Goulart${^*}$ and Guenther Rehm${^{**}}$
\thanks{$^*$Corresponding author: {\tt\footnotesize{idris.kempf@eng.ox.ac.uk}}. The authors are with the Department of Engineering Science, University of Oxford, Oxford, UK. This research is supported by the Engineering and Physical Sciences Research Council (EPSRC) with a Diamond CASE studentship.}
\thanks{${^{**}}$Diamond Light Source, Didcot, UK.}
}

\maketitle

\begin{abstract}
We introduce a novel structured controller design for the electron beam stabilization problem of the UK's national synchrotron light source. Because changes to the synchrotron will not allow the application of existing control approaches, we develop a novel method to diagonalize the multi-input multi-output (MIMO) system. A generalized singular value decomposition (GSVD) is used to simultaneously diagonalize the actuator response matrices, which is applicable to an arbitrary number of actuator dynamics in a cross-directional setting. The resulting decoupled systems are regulated using mid-ranged control and the controller gains derived as a function of the generalized singular values. In addition, we exploit the inherent block-circulant symmetry of the system. The performance of our controller is demonstrated using simulations that involve machine data.
\end{abstract}


\section{Introduction}\label{sec:introduction}
Diamond Light Source is the UK's national synchrotron facility that produces synchrotron light for research. This exceptionally bright synchrotron light is used as a source for various microscopic experiments, particularly in the X-ray region of the electromagnetic spectrum \cite{SYNCRAD}. The light is produced by electrons that circulate around a 560m storage ring, which have been previously accelerated to 3GeV. An assembly of magnets produces a strong magnetic field that confines the electrons in the storage ring. Large magnet arrays steer and focus the electron beam whilst smaller corrector magnets attenuate the vibrations induced by disturbances. These disturbances arise from vibrations transmitted by the girders on which the magnet arrays are attached, or from internal devices, such as the insertion devices used to extract the synchrotron light. While the large magnet arrays are actuated with feedforward control, the correctors are controlled by a feedback loop. The deviation of the particles from their ideal orbital path is measured by beam position monitors and the stabilizing set-points are routed back as inputs to the corrector magnets. At Diamond, this feedback reduces the trajectory error of the electrons to 5$\mu$m in the horizontal and 600nm in the vertical plane \cite{PHDSANDIRA}. This trajectory error must be minimized in order to retain certain properties of the synchrotron light.

Diamond Light Source has completed the conceptual design phase of a significant upgrade (Diamond-II), which will increase the brightness of the synchrotron light by raising the electron beam energy from 3GeV to 3.5GeV \cite{DIAMONDII}. Currently only one type of corrector magnet is used \cite{SANDIRACONTROLDESIGN}, but the upgrade will instead use two types, for both high and low bandwidth corrections. The number of sensors and actuators will be increased from 172 to 252 and 173 to 396, respectively. In addition, the sampling frequency will be increased from 10kHz to 100kHz. As a consequence of these changes, the feedback controller for the corrector magnets must be re-designed.

The design and analysis of electron beam stabilization controllers is a mature subject (with contributions from \cite{BOGEOLD, FOFBSVD, SANDIRACONTROLDESIGN, SANDIRAOPTIMAL, PAROBUST, SANDIRAMULTIARRAY}) and enjoys many parallels to the controller design for cross-directional control \cite{AMMAR2015283, PHDSANDIRA, SANDIRAMULTIARRAY}. The term cross-directional refers to systems for which the spatial and dynamic components of the actuator response can be separated \cite{DUNCANUMIST}. The vast majority of these studies focus on the case when one type of actuator is used (exceptions are  \cite{SANDIRAMULTIARRAY} or \cite{PHDFAN}). The common approach for the controller design in this case is to diagonalize the MIMO system using a singular value decomposition (SVD). The system is then said to be projected into modal space, where each mode is associated with a singular value (SV) and the modes are regulated independently using single-input single-output (SISO) controllers. The introduction of multiple types of actuators prohibits the former decoupling approach \cite{SANDIRAMULTIARRAY}; because the controllable subspaces of slow and fast actuators overlap, the modal decomposition does not lead to a diagonalized system and not all modes can be controlled independently. A heuristic approach to decouple the MIMO system is presented in \cite{SANDIRAMULTIARRAY}. Based on the principal angles between these subspaces, a decoupling matrix is derived which transforms the original system into a set of SISO and two-input single-output (TISO) systems. This approach, however, is not applicable for Diamond-II, because the subspace generated by the fast actuators is entirely contained in the subspace generated by slow actuators, so the analysis of the principal angles becomes redundant and the use of heuristics unavoidable.

This paper presents an alternative approach that uses the generalized singular value decomposition (GSVD) \cite{PAIGE_GSVD} to simultaneously diagonalize the systems formed by slow and fast actuators, resulting in two sets of SISO and TISO systems. By contrast to \cite{SANDIRAMULTIARRAY, duncan2008a}, this approach does not require any heuristics to derive a decoupling matrix. Moreover, our approach is applicable to an arbitrary number of actuator arrays \cite{HOGSVD}. In addition, we exploit the inherent block-circulant structure of the control problem~\cite{SYMSYNCHARXIV} and project the system dynamics into the spatial Fourier domain. The circulant symmetry has been shown to be key to the robustness analysis of electron beam stabilization controllers \cite{PAROBUST} and provides a significant potential in increasing the computational efficiency \cite{BCINVADMMCONF}.

\textit{Notation and Definitions} Let $\otimes$ denote the Kronecker product and $\I_n$ represent the identity matrix in $\R^{n\times n}$.  For a scalar, vector or matrix $a$, let $\bar{a}$ denote its complex conjugate; Let $\trans{a}$ denote its transpose and $\herm{a}$ its Hermitian transpose.
\section{Preliminaries}\label{sec:preliminaries}
\subsection{Process Model}\label{sec:pm}
The Diamond storage ring is composed out of $n=6$ identical sections. The relationship between the beam displacements $\inRv{y(t)}{N_y}$ measured at $N_y\!=\!n n_y\!=\!252$ locations around the ring and the $N_u\!=\!n n_u\!=\!396$ corrector magnet inputs $\inRv{u(t)}{N_u}$ at time $t=k \tau$, where $\tau$ is the sampling time, is given by
\begin{align}
y[k] = P(\inv{z})u[k] + d[k],\label{eq:initialmodel}
\end{align}
where $\inv{z}$ represents the backward shift operator and $y[k]=y(k\tau)$. The transfer function $P(\inv{z})$ takes the form
\begin{align}
P(\inv{z}) = R (\I_n\otimes G(\inv{z})),\label{eq:tfmatrix}
\end{align}
with $\inR{R}{N_y}{N_u}$ denoting the orbit response matrix and $\inC{G(\inv{z})}{n_u}{n_u}$ the diagonal actuator response matrix, which is repeated $n$ times around the storage ring. The storage ring features two different kinds of actuators; $N_s\!=\!nn_s\!=\!252$ slow magnets with a bandwidth $a_s = 80\text{ rad}\inv{s}$ but strong magnetic field, and $N_f\!=\!nn_f\!=\!144$ fast magnets with a bandwidth $a_f = 12\text{ krad}\inv{s}$ but weak magnetic field. Each actuator is modeled as a first-order system subjected to a delay of $\mu=7$ time steps and the diagonal elements of $G(\inv{z})$ take the form
\begin{align}\label{eq:actuatormodel}
g_{(\cdot)}(\inv{z})= z^{\sm(\mu+1)} \frac{1-e^{\sm a_{(\cdot)}\tau}}{1-\inv{z}e^{\sm a_{(\cdot)}\tau}},
\end{align}
where $(\cdot)=\{s,f\}$ depending on the pattern of actuators. By grouping the columns corresponding to $g_s(\inv{z})$ and $g_f(\inv{z})$, system \eqref{eq:initialmodel} can be represented as
\begin{align}\label{eq:fastslowmodel}
y[k] = R_s g_s(\inv{z}) u_s[k] + R_f g_f(\inv{z}) u_f[k] + d[k].
\end{align}
In practice, the power supplies of the magnets are subject to slew-rate and amplitude constraints. While this paper focuses on the decomposition, the following control algorithm can be robustly extended with non-linear blocks to consider actuator constraints~\cite{SANDIRAWINDUP}.

\subsection{Block-Circulant Matrices}\label{sec:bcmatrices}
The placement of sensors and actuators determines the structure of the orbit response matrix $R$. If the arrangement is repeated $n$ times around the ring, the matrix $R$ inherits an $n$-fold circulant symmetry. Let $\BC(n,p,m)$ denote the set of real- or complex-valued block-circulant matrices of the form
\begin{align}
B=\CircBB,\qquad \inC{b_i}{p}{m}.\label{eq:bc}
\end{align}
The circulant structure entails that $B\in\BC(n,p,m)$ iff $\hat{B}$ is block-diagonal \cite{CIRCBOOK}:
\begin{align}
\hat{B} = (F_n^*\otimes\I_p) B (F_n\otimes\I_m) = \begin{bmatrix}
\beta_0\\ & \ddots\\ & & \beta_{n-1}
\end{bmatrix},\label{eq:bdiagB}
\end{align}
where $\inC{\beta_i}{p}{m}$. For real-valued $B$, the blocks $\beta_i$ possess additional structure: If $n$ is even, $\beta_0$ and $\beta_{n/2}$ are real while for $i=1,\dots,n/2\!\sm\!1$ it holds that $\beta_i=\bar{\beta}_{n-i}$. If $n$ is odd, the only real-valued block is $\beta_0$ and the latter holds for $i=1,\dots,(n\sm1)/2$. The matrix $F_n$ with $F_n^*F_n=\I_n$ is called the Fourier matrix and  the product $F_nx$ yields the coefficients of the discrete Fourier transformation of the vector $x$. Because the Fourier matrix appears in \eqref{eq:bdiagB}, the computation speed of a matrix-vector multiplication $Bx$ can be significantly increased by transforming into the Fourier domain, i.e. by computing
\begin{subequations}\begin{align}
Bx &= \trans{\pi}_p \left(\I_p\otimes F_n^*\right) \pi_p \hat{B} \trans{\pi_m} \left(\I_m\otimes F_n\right) \pi_m x,\\
&=: \mathcal{F}^{- 1}(\hat{B}\mathcal{F}(x))\label{eq:FFT}
\end{align}\end{subequations}
where $\pi_p$ and $\pi_m$ are unitary permutation matrices associated to the reversion of the Kronecker products \cite{ROSE}. The computational efficiency arises from the possibility of using $m$ parallel FFTs for computing products such as $\trans{\pi_m} (\I_m\otimes F_n) \pi_m x$ and the fact that $\hat{B}$ is block-diagonal and partly redundant for real-valued $B$. Assuming that $B$ is dense, the number of cells $n$ a power of $2$ and neglecting the complex arithmetic as well as the redundant blocks of $\hat{B}$, the computation time is reduced by a factor
\begin{align}
\frac{\co{\mathcal{F}^{- 1}(\hat{B}\mathcal{F}(x))}}{\co{Bx}}=
\frac{(m+p)n\log_2 n + nmp}{n^2 mp}.\label{eq:speedup}
\end{align}
The Diamond storage ring features a $6$-fold circulant symmetry with blocks of size $42\times 66$. With an implementation that considers the redundant blocks in~\eqref{eq:FFT}, we measured that one can achieve a reduction in computation time between $70\%$ and $80\%$ \cite{BCINVADMMCONF,SYMSYNCHARXIV}.

\subsection{Generalized Singular Value Decomposition}\label{sec:gsvd}
The generalized singular value decomposition (GSVD) simultaneously diagonalizes two matrices. The following theorem has been transposed and adapted to the given problem dimensions.
\begin{theorem}[GSVD {\cite{GOLUB3}[Thm. 8.7.4]}]\label{thm:gsvd}
Given $\inC{A}{q}{q}$ and $\inC{B}{q}{m}$ with $m < q$, there exist orthogonal $\inC{U_A}{q}{q}$ and $\inC{U_B}{m}{m}$ and a nonsingular $\inC{X}{q}{q}$ such that
\begin{align}\label{eq:gsvddecomp}
A = X \begin{bmatrix}
S_A & 0\\ 0& \I_{q-m}
\end{bmatrix}\herm{U_A},\qquad
B = X \begin{bmatrix}
S_B \\ 0
\end{bmatrix}\herm{U_B},
\end{align}
where the diagonal matrices $S_A,\inR{S_B}{m}{m}\succ 0$ are referred to as generalized singular values and satisfy
\begin{align}
\trans{S_A}S_A + \trans{S_B}S_B = \I_m.\\\nonumber
\end{align}
\end{theorem}
As pointed out in \cite{PAIGE_GSVD}, the GSVD of matrices $A$ and $B$ is essentially the SVD of the concatenated matrices, i.e.
\begin{align}\label{eq:largesvd}
C=\begin{bmatrix}\herm{A} \\ \herm{B}\end{bmatrix} = Q \begin{bmatrix}\Sigma\\ 0 \end{bmatrix} \herm{\Psi} = \begin{bmatrix}
Q_{11}&\cdot_{\phantom{21}}\\Q_{21}& \cdot_{\phantom{22}}\end{bmatrix} \begin{bmatrix}\Sigma \\ 0\end{bmatrix} \herm{\Psi},
\end{align}
where $\inC{Q_{11}}{q}{q}$, $\inC{Q_{21}}{m}{q}$ and $\inR{\Sigma}{q}{q}$ is diagonal. We assume that $\rank(C)=q$ and therefore $\Sigma \succ 0$. Note, however, that this is not a requirement on the invertibility of $X$. If $\rank(C)<q$, the columns of $Q$ in \eqref{eq:largesvd} corresponding to the zero SVs are removed and the subsequent computations carried out using the positive set of SVs only. Because the matrix $\herm{[\herm{Q_{11}},\herm{Q_{21}}]}$ has orthonormal columns, a CS decomposition \cite[Thm. 2.5.2]{GOLUB3} yields 
\begin{align}\label{eq:csdecomp}
Q_{11}=U_A\begin{bmatrix}
S_A & 0\\ 0& \I_{q\sm m}
\end{bmatrix}
\herm{V}, \quad Q_{21}=U_B\left[S_B\,\,0\right]\herm{V}.
\end{align}
Theorem \ref{thm:gsvd} follows from combining \eqref{eq:largesvd} with \eqref{eq:csdecomp} and from defining the invertible matrix
\begin{align}
\herm{X}=\herm{V}\Sigma \herm{\Psi}.
\end{align}
It is important to note that the singular values of the concatenated matrix $[A\,\,\,B]$ are contained in $X$, i.e. the shared left-hand side of decomposition \eqref{eq:gsvddecomp}. Clearly, if $[A\,\,\,B]$ is ill-conditioned, then $X$ is ill-conditioned and regularizing the inverse of the concatenated matrix is the same as regularizing the inverse of $X$.

Recent research showed that Theorem \ref{thm:gsvd} can be extended to more than two matrices \cite{HOGSVD}. The SVs of the concatenated matrix remain in the matrix $X$ shared among the individual decompositions. The method presented in this paper could therefore be extended to an arbitrary number of actuators with different dynamics.
\section{System Decomposition}\label{sec:sd}
\subsection{Fourier Decomposition}\label{sec:fd}
Applying the first change of coordinates,
\begin{gather}\begin{aligned}\label{eq:cos}
\hat{y}&=(F_n^*\otimes \I_{n_y})y_,\qquad\,\hat{d}=(F_n^*\otimes \I_{n_y})d,\\
\hat{u}_s&=(F_n^*\otimes \I_{n_s})u_s,\quad\hat{u}_f=(F_n^*\otimes \I_{n_f})u_f,
\end{aligned}\end{gather}
maps \eqref{eq:fastslowmodel} into the spatial Fourier domain:
\begin{align}\label{eq:fastslowmodelfourier}
\hat{y} = \hat{R}_s g_s(\inv{z}) \hat{u}_s + \hat{R}_f g_f(\inv{z}) \hat{u}_f + \hat{d}.
\end{align}
Since $R_{s},R_{f}$ are block-circulant, $\hat{R}_{s}$ and $\hat{R}_{f}$ are block-diagonal. For the nominal controller design, incorporating transformation \eqref{eq:cos} merely represents a coordinate transformation. In contrast to the circulant case, i.e. when the blocks $b_i$ in \eqref{eq:bc} are scalars, \eqref{eq:cos} does not yield a set of decoupled SISO systems but a set of decoupled MIMO systems \eqref{eq:fastslowmodelfourier}. 

\begin{figure*}[htpb]
\centering
\begin{tikzpicture}[align=center,node distance=0.4cm,
	    every node/.style={inner sep=2pt,rectangle, minimum height=1.7em, text centered},
	    block/.style={draw,inner sep=2pt},
	    sum/.style={draw,minimum height = 4pt,circle,inner sep=1pt},
	    nnode/.style={draw,fill=black,minimum height=2pt,circle,inner sep=0pt}]
	    
	\node[] (y11) [] {};
	\node[sum] (y12) [below of=y11] {};
	\node[] (y13) [below of=y12] {};
	\node[] (y10) [above of=y11] {};
	\node[] (y11add) [right of=y11] {};
	\node[] (y21) [right = \lonea of y11add] {};
	\node[block] (y22) [below of=y21] {\st{$C(z^{\sm 1})$}};
	\node[] (y23) [below of=y22] {};
	\node[] (y21add) [right of=y21] {};
	\node[] (y31) [right = \ltwo of y21add] {};
	\node[block] (y32) [below of=y31] {\st{$\bar{P}(z^{\sm 1})$}};
	\node[] (y41) [right = \ltwo of y31] {};
	\node[sum] (y42) [below of=y41] {};
	\node[] (y51) [right = \lone of y41] {};
	\node[nnode] (y52) [below of=y51] {};
	\node[] (y53) [below of=y52] {};
	\node[] (y61) [right = of y51] {};
	\node[] (y62) [below of = y61] {};
	\node[] (y71) [right = of y61] {};
	\node[] (y70) [above of = y71] {};
	
	\draw[-Latex, line width=\lw] (y12.east) -- (y22.west);
	\draw[-Latex, line width=\lw] (y22.east) -- (y32.west);
	\draw[-Latex, line width=\lw] (y32.east) -- (y42.west);
	\draw[line width=\lw] (y42.east) -- (y52.west);
	\draw[-Latex, line width=\lw] (y52.east) -- (y62.east);
	\draw[-Latex, line width=\lw] (y41.north) -- (y42.north);
	\draw[line width=\lw] (y52.south) -- (y53.center);
	\draw[-Latex, line width=\lw] (y53.center) -| (y12.south);
	
  	\node[] () [below right=\sshift and \ssshift of y12] {\ssm};
  	\node[] () [below left=\sshift and \ssshift of y42] {\ssmp};
  	\node[] () [above right=\sshift and \ssshift of y42] {\ssmp};

  	\node[] () [right = \lnshift of y62] {\st{$y[k]$}};
  	\node[] () [yshift=0.1cm][right = \lnshift of y41] {\st{$d[k]$}};
  	\node[] (A) [left = \lone of y10] {\textbf{\large{A}}};
  	
	\node[] (z11) [right = \ltwo of y71] {};
	\node[sum] (z12) [below of=z11] {};
	\node[] (z11add) [right of=z11] {};
	\node[] (z21) [right = \lonea of z11add] {};
	\node[block] (z22) [below of=z21] {\st{$Q(z^{\sm 1})$}};
	\node[] (z31) [right = \ltwo of z21] {};
	\node[nnode] (z32) [below of=z31] {};
	\node[block] (z41) [right = \ltwo of z31] {\st{$\bar{P}(z^{\sm 1})$}};
	\node[] (z42) [below of=z41] {};
	\node[block] (z43) [below of = z42] {\st{$P(z^{\sm 1})$}};
	\node[sum] (z51) [right = \ltwo of z41] {};
	\node[] (z50) [above of = z51] {};
	\node[nnode] (z61) [right = of z51] {};
	\node[] (z62) [below of=z61] {};
	\node[sum] (z63) [below of=z62] {};
	\node[] (z64) [below of = z63] {};
	\node[] (z71) [right = of z61] {};
	\node[] (z72) [below of = z71] {};
	
	\draw[-Latex, line width=\lw] (z12.east) -- (z22.west);
	\draw[line width=\lw] (z22.east) -- (z32.west);
	\draw[-Latex, line width=\lw] (z32.center)  |- (z41.west);
	\draw[-Latex, line width=\lw] (z41.east) -- (z51.west);
	\draw[line width=\lw] (z51.east) -- (z61.west);
	\draw[-Latex, line width=\lw] (z61.east) -- (z71.east);
	\draw[-Latex, line width=\lw] (z50.north) -- (z51.north);
	\draw[-Latex, line width=\lw] (z32.south) |- (z43.west);
	\draw[-Latex, line width=\lw] (z43.east) -- (z63.west);
	\draw[-Latex, line width=\lw] (z61.south) -- (z63.north);
	\draw[line width=\lw] (z63.south) -- (z64.center);
	\draw[-Latex, line width=\lw] (z64.center) -| (z12.south);
	
  	\node[] () [below right=\sshift and \ssshift of z12] {\ssm};
  	\node[] () [below left=\sshift and \ssshift of z63] {\ssm};
  	\node[] () [below left=\sshift and \ssshift of z51] {\ssmp};
  	\node[] () [above right=\sshift and \ssshift of z51] {\ssmp};
  	\node[] () [above right=\sshift and \ssshift of z63] {\ssmp};

  	\node[] () [right = \lnshift of  z71] {\st{$y[k]$}};
  	\node[] () [yshift=0.1cm][right = \lnshift of z50] {\st{$d[k]$}};
  	\node[] () at (A -| z11) {\textbf{\large{B}}};

	\node[] (x11) [below = \ltwo of y13] {};
	\node[sum] (x12) [below of=x11] {};
	\node[] (x11add) [right of=x11] {};
	\node[] (x21) [right = \lonea of x11add] {};
	\node[block] (x22) [below of=x21] {\st{$\mathcal{F}$}};
	\node[] (x23) [below of=x22] {};
	\node[] (x31) [right = \lone of x21] {};
	\node[nnode] (x32) [below of=x31] {};
	\node[block] (x41) [right = \lone of x31] {\st{$X^\dagger_f$}};
	\node[] (x42) [below of=x41] {};
	\node[block] (x43) [below of=x42] {\st{$X^\dagger_s$}};
	\node[block] (x51) [right = \lone of x41] {\st{$K_f(z^{\sm 1})$}};
	\node[] (x52) [below of=x51] {};
	\node[block] (x53) [below of=x52] {\st{$K_s(z^{\sm 1})$}};
	\node[block] (x61) [right = \lone of x51] {\st{$U_f$}};
	\node[] (x62) [below of=x61] {};
	\node[block] (x63) [below of=x62] {\st{$U_s$}};
	\node[] (x71) [right = \lone of x61] {};
	\node[] (x72) [below of=x71] {};
	\node[] (x73) [below of=x72] {};
	\node[] (x81) [right = \ltwo of x71] {};
	\node[block] (x82) [below of=x81] {\st{$\mathcal{F}^{\sm 1}$}};
	\node[] (x83) [below of=x82] {};
	\node[] (x91) [right = \lone of x81] {};
	\node[nnode] (x92) [below of=x91] {};
	\node[] (x93) [below of=x92] {};
	\node[block] (x101) [right = \ltwo of x91] {\st{$\bar{P}(z^{\sm 1})$}};
	\node[] (x102) [below of=x101] {};
	\node[block] (x103) [below of=x102] {\st{$P(z^{\sm 1})$}};
	\node[sum] (x111) [right = \ltwo of x101] {};
	\node[] (x112) [below of=x111] {};
	\node[] (x113) [below of=x112] {};
	\node[] (x110) [above of=x111] {};
	\node[nnode] (x121) [right = of x111] {};
	\node[] (x122) [below of=x121] {};
	\node[sum] (x123) [below of=x122] {};
	\node[] (x124) [below of=x123] {};
	\node[] (x125) [below of=x124] {};
	\node[] (x131) [right = of x121] {};	
	
	\draw[-Latex, line width=\lw] (x12.east) -- (x22.west);
	\draw[line width=\lw] (x22.east) -- (x32.center);
	\draw[-Latex, line width=\lw] (x32.center) |- (x41.west);
	\draw[-Latex, line width=\lw] (x32.center) |- (x43.west);
	\draw[-Latex, line width=\lw] (x41.east) -- (x51.west);
	\draw[-Latex, line width=\lw] (x51.east) -- (x61.west);
	\draw[-Latex, line width=\lw] (x61.east) -- (x71.west);
	\draw[-Latex, line width=\lw] (x43.east) -- (x53.west);
	\draw[-Latex, line width=\lw] (x53.east) -- (x63.west);
	\draw[-Latex, line width=\lw] (x63.east) -- (x73.west);
	\draw[line width=\lw*3] (x73.south) -- (x71.north);
	\draw[-Latex, line width=\lw] (x72.center) -- (x82.west);
	\draw[line width=\lw] (x82.east) -- (x92.center);
	\draw[-Latex, line width=\lw] (x92.center) |- (x101.west);
	\draw[-Latex, line width=\lw] (x92.center) |- (x103.west);
	\draw[-Latex, line width=\lw] (x101.east) -- (x111.west);
	\draw[line width=\lw] (x111.east) -- (x121.west);
	\draw[-Latex, line width=\lw] (x121.east) -- (x131.east);
	\draw[-Latex, line width=\lw] (x110.north) -- (x111.north);
	\draw[-Latex, line width=\lw] (x121.center) -- (x123.north);
	\draw[-Latex, line width=\lw] (x103.east) -- (x123.west);
	\draw[line width=\lw] (x123.south) -- (x125.center);
	\draw[-Latex, line width=\lw] (x125.center) -| (x12.south);
	
  	\node[] () [below right=\sshift and \ssshift of x12] {\ssm};
  	\node[] () [above right=\sshift and \ssshift of x111] {\ssmp};
  	\node[] () [below left=\sshift and \ssshift of x111] {\ssmp};
  	\node[] () [below left=\sshift and \ssshift of x123] {\ssm};
  	\node[] () [above right=\sshift and \ssshift of x123] {\ssmp};
  	
  	\node[] () [right = \lnshift of x131] {\st{$y[k]$}};
  	\node[] () [yshift=0.1cm][right = \lnshift of x110] {\st{$d[k]$}};
  	\node[] () at (A |- x110) {\textbf{\large{C}}};
  	

\end{tikzpicture}
\caption{(\textbf{A}) Standard feedback and (\textbf{B}) IMC structure. (\textbf{C}) IMC structure for \eqref{eq:fastslowmodelmodal} including transformations \eqref{eq:cos} and \eqref{eq:cosmodal}.}\label{fig:feedback}
\end{figure*}
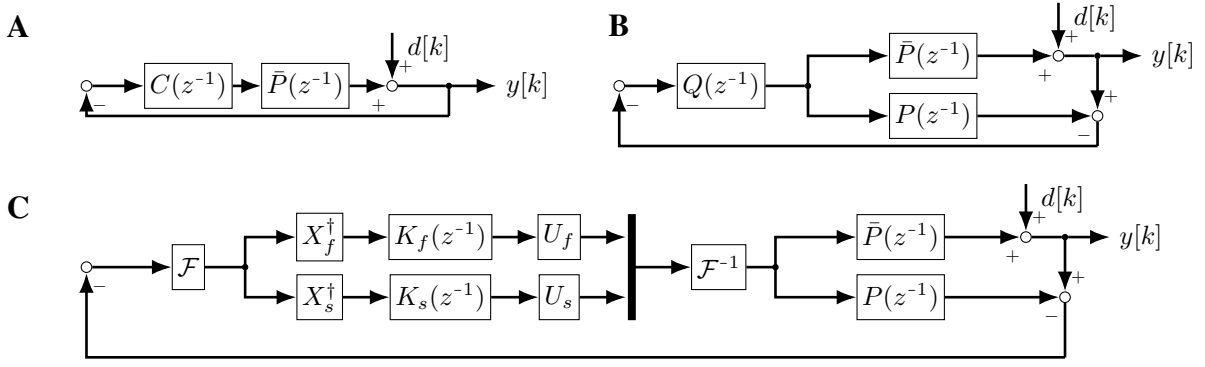
\subsection{General Modal Decomposition}\label{sec:gmd}
Next, apply the GSVD of Theorem \ref{thm:gsvd} to \eqref{eq:fastslowmodelfourier} or equivalently to each of the $n$ blocks separately,
\begin{align}\label{eq:blockdecomp}
\sigma_i = X_i \begin{bmatrix}
S_{s,i} & 0\\ 0& \I_{n_s-n_f}
\end{bmatrix}\herm{U_{s,i}},\quad
\phi_i = X_i \begin{bmatrix}
S_{f,i} \\ 0
\end{bmatrix}\herm{U_{f,i}},
\end{align}
with $\hat{R}_s\!=\!\diag(\sigma_0,\dots,\sigma_{n\sm 1})$ and $\hat{R}_f\!=\!\diag(\phi_1,\dots,\phi_{n\sm 1})$. Define $X,U_s,U_f,S_s$ and $S_f$ as the diagonal concatenation of the corresponding matrices in \eqref{eq:blockdecomp} for $i\!=\!0,\dots,n\sm 1$, where we include $\I_{n_s-n_f}$ in $S_s$ and the zeros in $S_f$. Applying the second change of coordinates,
\begin{align}\label{eq:cosmodal}
\tilde{y}=\inv{X}\hat{y},\,\,\,\tilde{d}=\inv{X}\hat{d},\,\,\,
\tilde{u}_s=\herm{U}_s\hat{u}_s,\,\,\,\tilde{u}_f=\herm{U}_f\hat{u}_f,
\end{align}
maps \eqref{eq:fastslowmodelfourier} into the generalized mode space of the Fourier coefficients:
\begin{align}\label{eq:fastslowmodelmodal}
\tilde{y} = S_s g_s(\inv{z}) \tilde{u}_s + S_f g_f(\inv{z}) \tilde{u}_f + \tilde{d},
\end{align}
where the diagonal $S_{s}$, $S_{f}$ satisfy $\trans{S_s}S_s + \trans{S_f}S_f=\I_{N_y}$. The matrices $S_{s}$, $S_{f}$  can be interpreted as a splitting between a slow and a fast input direction -- corresponding to two columns of matrices $U_{s}$, $U_{f}$ -- onto an output direction and amplification defined by a column of $X$ containing the SVs of the orbit response matrix $R$. It determines which modes of $R$ can be simultaneously controlled by slow and fast actuators and which modes can be controlled by fast actuators only. In addition, for the case that $\rank(R)<N_y$ the decomposition would also expose the uncontrollable modes.

\section{Controller}\label{sec:controller}
\subsection{Internal Model Control}\label{sec:imc}
The internal model control (IMC) structure is suitable for systems with delay \cite{MORARIIMC2} and follows previous controller designs for the  DLS synchrotron \cite{SANDIRACONTROLDESIGN, SANDIRAMULTIARRAY, SANDIRAWINDUP, PAROBUST}. The IMC feedback structure is shown in Figure \ref{fig:feedback}.B. In contrast to the standard feedback structure \ref{fig:feedback}.A, a model $P(\inv{z})$ is evaluated in parallel to the plant $\bar{P}(\inv{z})$ and the model output subtracted from the feedback signal. If the model is accurate, i.e. $P(\inv{z})\equiv\bar{P}(\inv{z})$, this structure has the advantage that the closed-loop system is stable provided that plant and controller are stable. Because according to~\eqref{eq:actuatormodel} the finite poles of~\eqref{eq:fastslowmodelmodal} lie within the unit circles, the system in consideration is stable. Moreover, because the model $P(\inv{z})$ incorporates the non-minimum phase (NMP) parts of the plant, such as delays, the limitations imposed by the NMP terms do not have to be actively considered. The standard controller can be obtained by setting $C(\inv{z})=\inv{(\I - Q(\inv{z})P(\inv{z}))}Q(\inv{z})$.

In the absence of modeling errors, the closed-loop transfer function from the disturbance $d[k]$ to the output $y[k]$ is:
\begin{align}\label{eq:climc}
y[k] = (\I - P(\inv{z})Q(\inv{z}))d[k].
\end{align}
A common choice \cite{MORARIIMC2} for the IMC controller $Q(\inv{z})$ is
\begin{align}\label{eq:ctrimc}
Q(\inv{z}) = \inv{P}(\inv{z})T(\inv{z}),
\end{align}
where $T(\inv{z})$ is a desired closed-loop transfer function that includes all NMP terms of $P(\inv{z})$ and this choice produces the control input $u[k]\!=\!- Q(\inv{z})y[k]$.

\subsection{Mid-Ranging Control}\label{sec:mrc}
The concept of mid-ranging control is to distribute the control action over a given bandwidth onto different actuators, each one of them assuming control over a portion of the bandwidth. A detailed design procedure for the combination of mid-ranging control and IMC is given in \cite{SANDIRAHEATH}. For the TISO case, \eqref{eq:climc} is rewritten as
\begin{align}\label{eq:clmrc}
y[k] = \left[\I \!-\! P_s(\inv{z})Q_s(\inv{z}) \!-\! P_f(\inv{z})Q_f(\inv{z})\right]d[k].
\end{align}
The design procedure is as follows. Firstly, a fast transfer function $T_f(\inv{z})\!:=\!P_s(\inv{z})Q_s(\inv{z})+P_f(\inv{z})Q_f(\inv{z})$ is chosen which spans the desired bandwidth. Secondly, a transfer function $T_s(\inv{z})\!:=\!P_s(\inv{z})Q_s(\inv{z})$ is defined with a slow bandwidth. Transfer functions $T_{s}(\inv{z})$, $T_{f}(\inv{z})$ must include the NMP parts of $P_{s}(\inv{z})$, $P_{f}(\inv{z})$. Finally, the controllers are obtained from
\begin{subequations}\label{eq:ctrmrc}\begin{align}
Q_s(\inv{z}) &= \inv{P}_s(\inv{z})T_s(\inv{z})\\
\quad Q_f(\inv{z}) &= \inv{P}_f(\inv{z})\left(T_f(\inv{z})-T_s(\inv{z})\right). 
\end{align}\end{subequations}
Choosing the controllers as in \eqref{eq:ctrmrc} results in the slow  and fast actuators regulating the lower and higher frequency content of $d[k]$, respectively.

\subsection{Controller Design: Dynamical Part}\label{sec:dynpar}
The MIMO system \eqref{eq:fastslowmodelmodal} solely comprises diagonal matrices and reduces the problem to $n_f$ TISO and $n_s \sm n_f$ SISO systems for each spatial Fourier frequency $i=1,\dots,n$:
\begin{align}\label{eq:TISOSISO}
\tilde{y}^{i} = \begin{bmatrix}
S_{s,i} & 0\\ 0& \I
\end{bmatrix}g_s(\inv{z}) \tilde{u}_{s}^{i}+
\begin{bmatrix}
S_{f,i} \\ 0
\end{bmatrix}g_f(\inv{z}) \tilde{u}_{f}^{i} + \tilde{d}^i,
\end{align}
where $\I =\I_{n_s\sm n_f}$. For each spatial frequency $i$, we will design $n_f$ identical TISO controllers $Q_s(\inv{z})$, $Q_f(\inv{z})$ and $n_s\sm n_f$ identical SISO controllers $Q(\inv{z})$. For the slow actuators, we will use the same controller for the TISO and the SISO parts, i.e. $Q_s(\inv{z})=Q(\inv{z})$. 

The transfer functions from Sections \ref{sec:imc} and \ref{sec:mrc}, $P(\inv{z}), P_s(\inv{z})$ and $P_f(\inv{z})$, are identified as $g_s(\inv{z}), g_s(\inv{z})$ and $g_f(\inv{z})$, respectively. The closed-loop transfer functions include the delay and are chosen to be identical for each generalized mode:
\begin{align}
T_{(\cdot)}(\inv{z})&=z^{\sm(\mu+1)}\frac{1-e^{-\lambda_{(\cdot)} \tau}}{1-\inv{z}e^{-\lambda_{(\cdot)} \tau}}
\end{align}
where $(\cdot)=\lbrace s,f\rbrace$ and with $\lambda_f\!=\! 1400$ Hz and $\lambda_s\!=\! 100$ Hz. $Q_{s}(\inv{z})$ and $Q_{f}(\inv{z})$ are obtained from~\eqref{eq:ctrmrc}. Because for the SISO controllers we choose $T(\inv{z})=T_{s}(\inv{z})$, it holds that $Q(\inv{z})=Q_s(\inv{z})$. The controller matrices for the entire system \eqref{eq:fastslowmodelmodal} are recovered as
\begin{subequations}\begin{align}
K_s(\inv{z})\!&=\!\I_{nn_s}Q_s(\inv{z})\label{eq:slowctr}\\
K_f(\inv{z})&= \I_n\!\otimes\!\begin{bmatrix}
\I_{n_f}\otimes Q_f(\inv{z})& 0 \\ 0 & 0
\end{bmatrix}.\label{eq:fastctr}
\end{align}\end{subequations}
Note that we included the zeros in the definition of $S_f$ in~\eqref{eq:fastslowmodelmodal} and therefore also include the zeros in~\eqref{eq:fastctr}.

\subsection{Controller Design: Regularization}\label{sec:regularization}
Because the orbit response matrix $R$ in \eqref{eq:tfmatrix} is ill-conditioned, inverting its SVs leads to relatively large controller gains in the directions associated with the higher-order modes \cite{SANDIRAOPTIMAL}. A common approach \cite{SANDIRACONTROLDESIGN, SANDIRAMULTIARRAY} to select the gains is to use Tikhonov regularization \cite{GOLUBTIKHONOV}, which selects the input $u[k]$ that for $z=1$ minimizes 
\begin{align}\label{eq:reg1}
\min_{u[k]} \twonorm{y[k] - P(\inv{z}) u[k]}^2 +\mu \twonorm{u[k]}^2,
\end{align}
where the scalar regularization parameter $\mu > 0$ weighs the residual norm against the input norm and $P(1)\!=\!\I$. From the solution to \eqref{eq:reg1},
\begin{align}\label{eq:reg1sol}
u[k] = \inv{(\trans{R}R+\mu I)}\trans{R} y[k]\reqdef X^\dagger y[k],
\end{align}
and the SVD of $R$, it can be seen how $\mu$ damps the inversion of small-magnitude SVs.

For the GSVD the matrix $X$ is not an orthogonal matrix and hence the change of coordinates \eqref{eq:cosmodal} is not norm-preserving. Minimizing the norm of the generalized modes $\twonorm{\tilde{y}}=\twonorm{\inv{X}\hat{y}}$ is therefore not equivalent to minimizing $\twonorm{\hat{y}}$ or $\twonorm{y}$. In order to find the controller gain matrices $X^\dagger_{s}$, $X^\dagger_{f}$ for the present system, the transformations involving $X$ are initially inverted. For mid-ranged control and steady-state, $g_s = 1$ and $\tilde{u}_f = 0$ and \eqref{eq:reg1} evaluates to
\begin{align}\label{eq:regslow}
\min_{\tilde{u}_s}\twonorm{\hat{y} \sm X S_s\tilde{u}_s}^2 +\mu_s\twonorm{\tilde{u}_s}^2.
\end{align}
For higher frequencies, $g_s \approx 0$ but $g_f \approx 1$ and \eqref{eq:reg1} becomes
\begin{align}\label{eq:regfast}
\min_{\tilde{u}_f}\twonorm{\hat{y} \sm X S_f\tilde{u}_f}^2 +\mu_f\twonorm{\tilde{u}_f}^2.
\end{align}
Analogously to \eqref{eq:reg1sol}, the regularized inverses $X_{s,f}^\dagger$ for slow and fast actuators are obtained as
\begin{align}\label{regslowfastsol}
X_{(\cdot)}^\dagger = \inv{(\trans{S}_{(\cdot)}\herm{X}X S_{(\cdot)}+\mu_{(\cdot)} I)}\trans{S}_{(\cdot)}\herm{X},
\end{align}
where $(\cdot)=\{s,f\}$. The controller gains \eqref{regslowfastsol} are combined with the controller dynamics from Section \ref{sec:dynpar} to yield the feedback system depicted in \ref{fig:feedback}.C. Following previous controller designs~\cite{SANDIRACONTROLDESIGN}, a regularization parameter $\mu_{s}=1$ is chosen for the slow actuators. In order to consider that the fast correctors generally produce weaker magnetic fields than slow ones, a parameter $\mu_{f}=10$ is selected. Other approaches exist that derive $X^\dagger$ based on minimizing a robust performance criterion \cite{OPTIMALROY}.
\subsection{Numerical Study}
\begin{figure}
\begin{subfigure}[b]{.5\columnwidth}
  	\centering
	\includegraphics[scale=0.9]{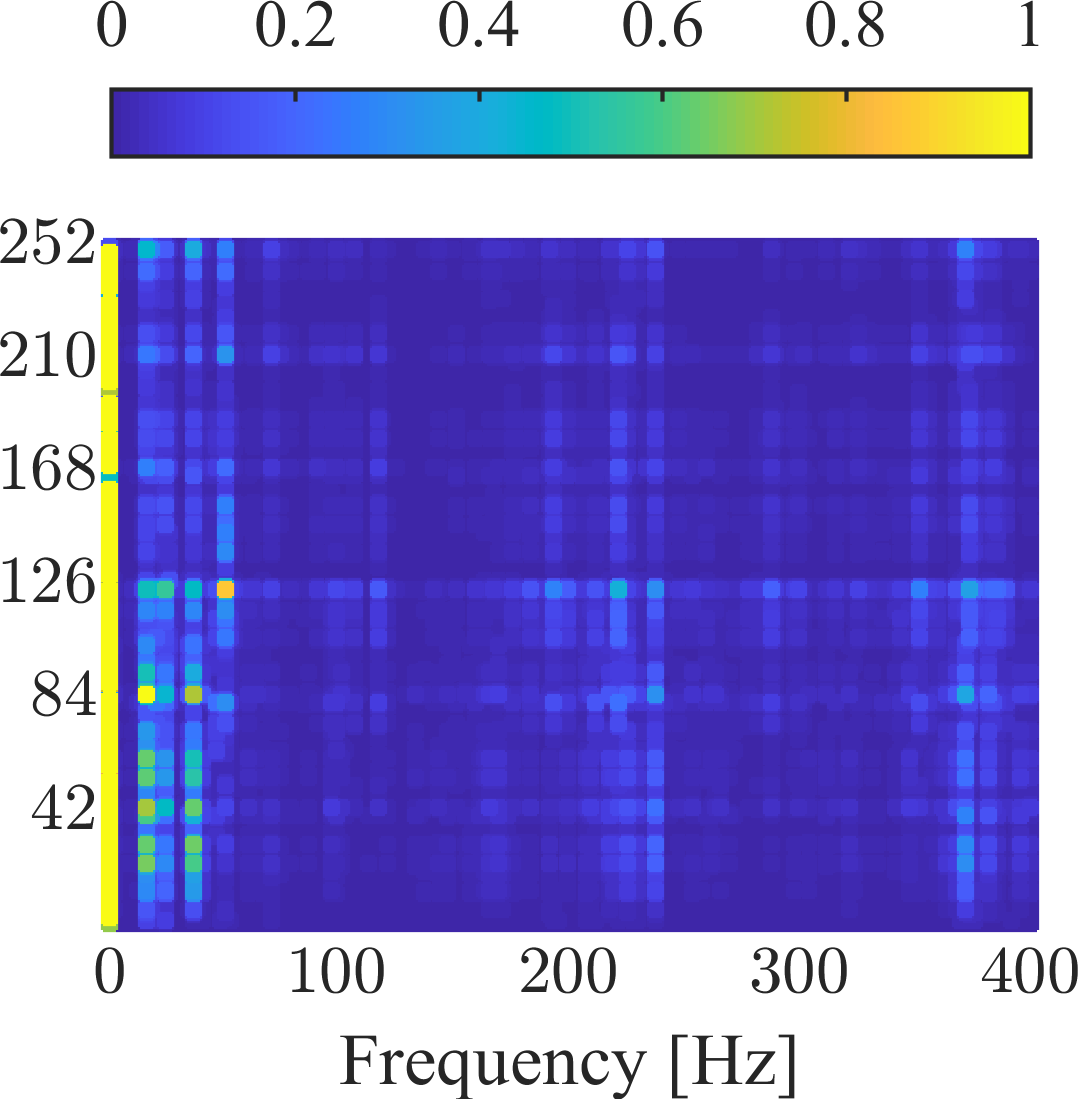}
  	\caption{Disturbances}
  	\label{fig:dist}
\end{subfigure}
\hfill%
\begin{subfigure}[b]{.48\columnwidth}
  	\centering
	\includegraphics[scale=0.9]{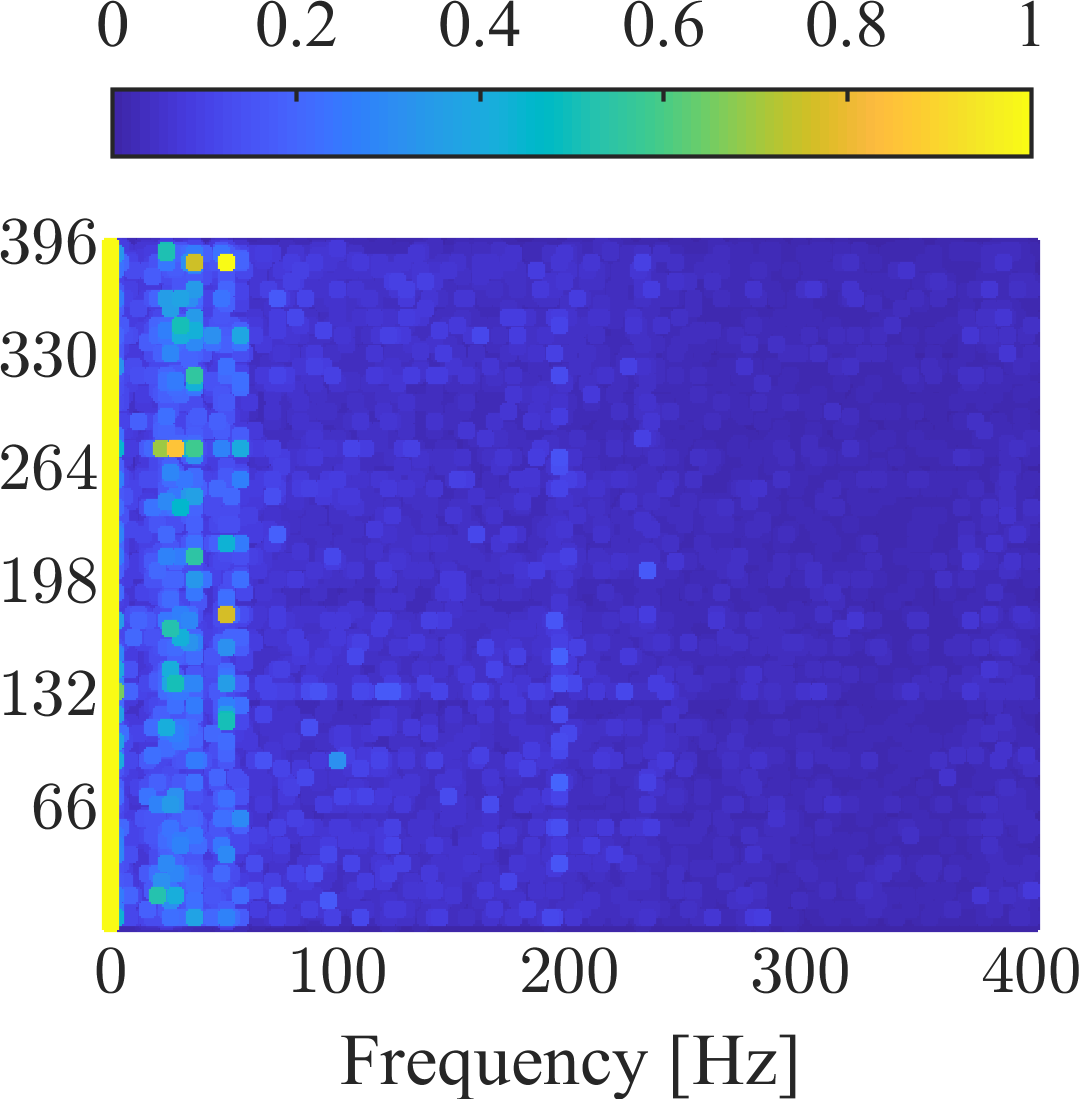}
  	\caption{Hypothetical actuators}
  	\label{fig:medium}
\end{subfigure}
\begin{subfigure}[b]{.5\columnwidth}
  	\centering
	\includegraphics[scale=0.9]{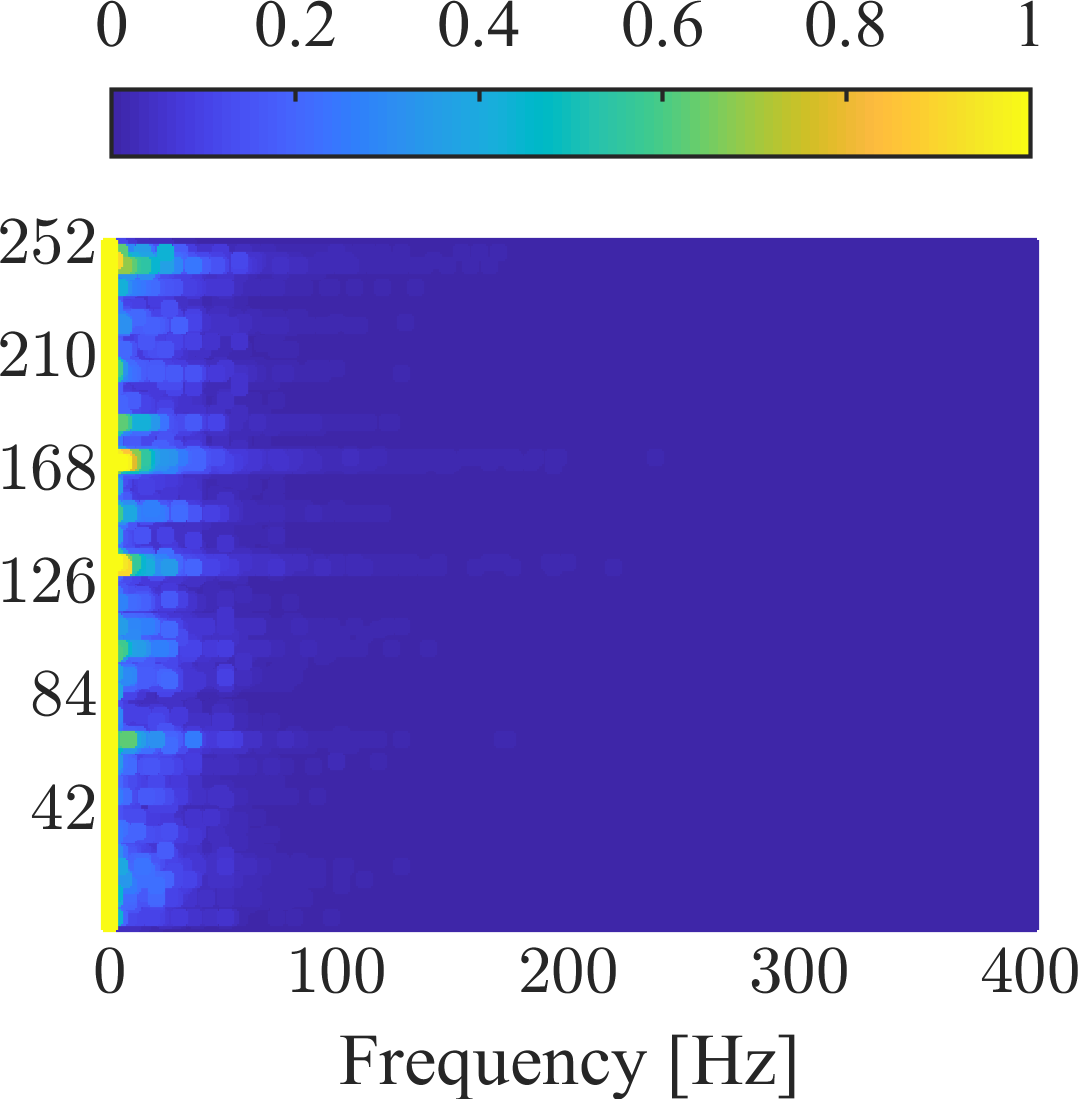}
  	\caption{Slow actuators}
  	\label{fig:slow}
\end{subfigure}
\hfill%
\begin{subfigure}[b]{.48\columnwidth}
  	\centering
	\includegraphics[scale=0.9]{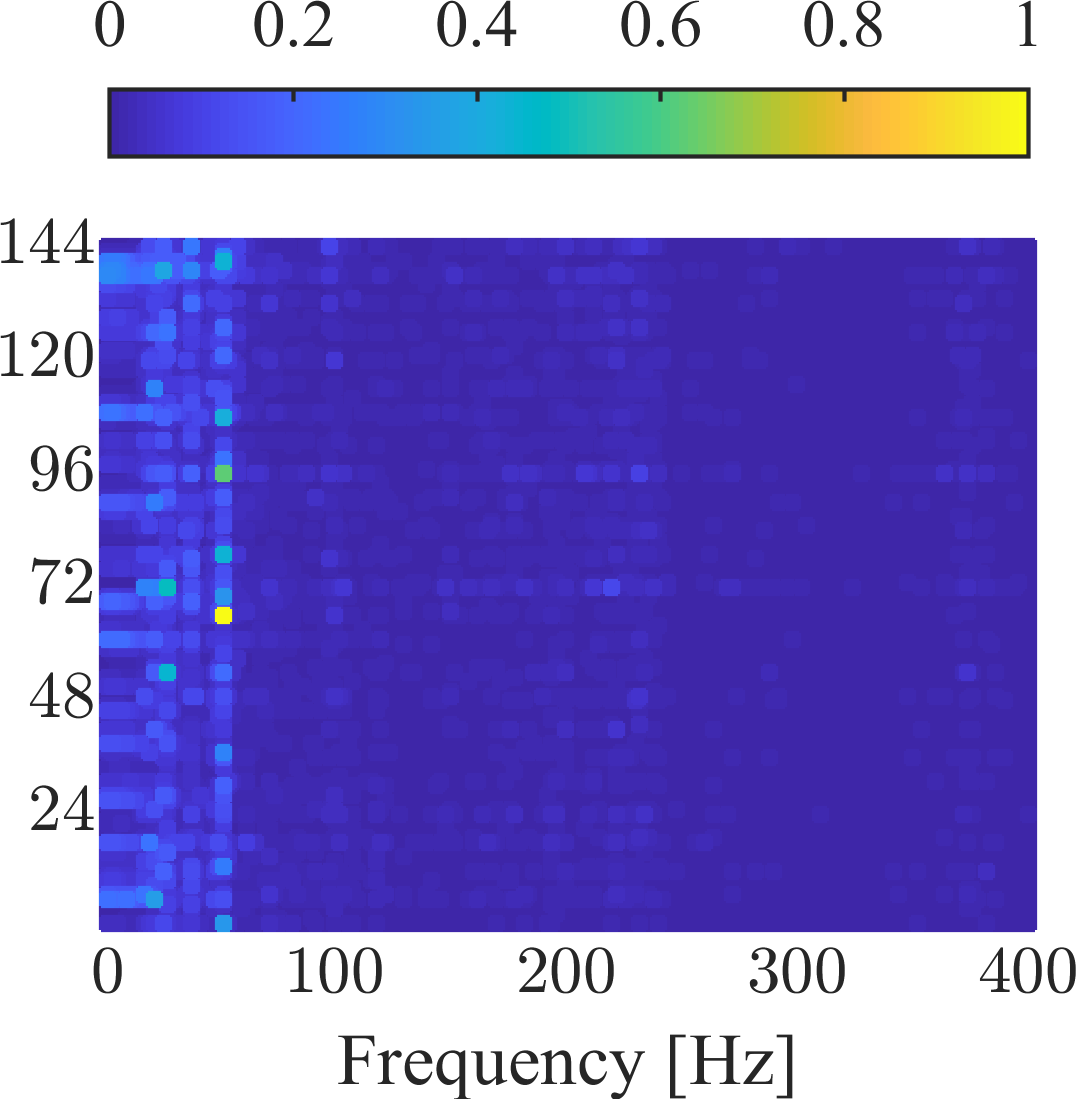}
  	\caption{Fast actuators}
  	\label{fig:fast}
\end{subfigure}
\caption{Magnitude plot of the power spectra of the $N_y=252$ augmented disturbances (a), $N_s\!+\!N_f=396$ hypothetical actuator inputs (b), $N_s=252$ slow (c) and $N_f=144$ fast actuators (d). All plots are normalized by their maximum value.}\label{fig:spectra}
\end{figure}

The simulation of the model~\eqref{eq:initialmodel} requires the disturbances $d(\inv{z})$ as an input. Because no such measurements are available for Diamond-II yet, the controller was simulated using modified $10$ kHz measurement data from the current storage ring, which has $172$ beam position monitors. The disturbance vector was augmented to fit the dimensions of Diamond-II, which has $252$ monitors. First, $80=252-172$ outputs were copied and appended to the measurements. Second, the augmented measurements were transformed into mode-space using an SVD of the Diamond-II orbit response matrix and the power spectrum of the modes was plotted, such as in~\cite[Chapter 3.5]{PHDSANDIRA}. Third, the disturbance spectrum was scaled in order to obtain a spectrum comparable to that in~\cite[Chapter 3.5]{PHDSANDIRA}, where the modes associated to large-magnitude singular values show a larger amplitude. Finally, the disturbances were transformed back into the original space using the modified power spectrum. The resulting power spectra  of the disturbances are shown in Figure~\ref{fig:dist}. The developed controller for system~\eqref{eq:fastslowmodel} with $N_s$ slow and $N_f$ fast actuators is compared against a hypothetical storage ring featuring $N_s\!+\!N_f$ fast actuators only. The hypothetical IMC system uses the same regularization parameter ($\mu=1$) and the same controller bandwidth as the TISO controller.

The power spectra of slow, fast and hypothetical actuators normalized by the corresponding maximum Fourier coefficient amplitudes are depicted in Figure~\ref{fig:medium}-\ref{fig:fast}. The magnitude plots show how the control effort of the hypothetical actuators (b) is split between slow (c) and fast actuators (d). While the hypothetical system is required to produce a large control effort for steady-state as well as a smaller control effort for frequencies up to $200$ Hz, the control action of slow actuators is limited to approximately 50Hz and the fast actuators mainly focus on higher frequencies and do not contribute to steady-state control action at all. In time domain, the hypothetical actuators have an average amplitude of $\abs{u[k]}=0.15$, while the slow and fast actuators have average amplitudes of $\abs{u_s[k]}=0.3$ and $\abs{u_f[k]}=0.01$, respectively. Even though actuator constraints have not actively been considered, our results suggest that the control system is likely to work in practice when the slow correctors are subjected to slew-rate and the fast ones subjected to amplitude constraints, respectively.

\begin{figure}[h]
\centering
\begin{overpic}[scale=1]{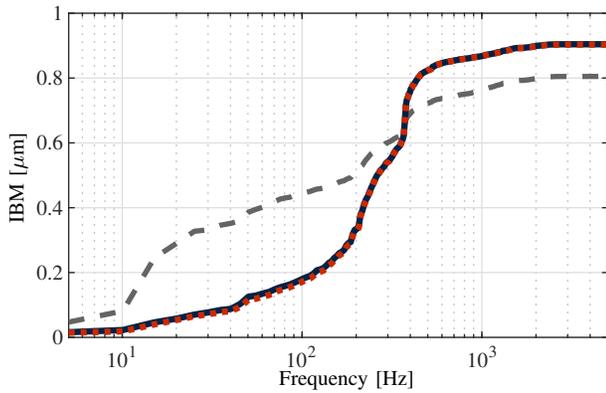}
\put(-4,26){\rotatebox{90}{\footnotesize{IBM [$\mu$m]}}}
\put(42,-1.5){\footnotesize{Frequency [Hz]}}
\end{overpic}
\caption{Integral beam motion (IBM) at the \nth{4} sensor for the uncontrolled beam (gray, dashed), the hyptothetical controller with $N_s+N_f$ fast actuators (red, dotted) and the presented controller (blue, solid).}\label{fig:ibm}
\end{figure}

The performance of the controller is measured using the integrated beam motion (IBM) -- the integrated square-root of the power spectral density. The IBM for a particular sensor can be expressed using its closed-loop sensitivity function $S(\inv{z})$ with $z=e^{i\omega T_s}$ as
$IBM(\omega) = \int_0^\omega \abs{S(\bar{\omega})d(\bar{\omega})}d\bar{\omega}$.
Figure \ref{fig:ibm} depicts the IBM for the hypothetical scenario (red) and for the system with slow and fast actuators (blue) at the \nth{4} sensor. The uncontrolled beam motion is represented by the dashed line (gray). Because the present and the hypothetical controllers have been set to cover the same bandwidth and use the same regularization parameter, the resulting IBM is almost identical for both cases.


\section{Conclusions}
\balance
This paper has presented the design of an electron beam stabilization controller for a synchrotron storage ring equipped with slow and fast corrector magnets. First, we demonstrated how the circulant symmetry of the accelerator can be exploited in order to increase the computing efficiency and block-diagonalize the feedback system. Second, we applied a GSVD in order to simultaneously diagonalize the dynamics of slow and fast actuators. The GSVD determines the modes of the orbit response matrix that can be jointly controlled by slow and fast actuators and those which cannot. Finally, we embedded the decomposed and diagonalized system in a mid-ranged IMC structure. The developed control system was simulated and compared against a hypothetical synchrotron storage ring equipped with high-bandwidth actuators. The control system yielded an identical disturbance attenuation while splitting the control effort between slow and fast actuators.

Although actuator constraints, such as slew-rate and amplitude limits, have not been considered in this paper, it is expected that including the constraints will be particularly relevant when slow and fast corrector magnets are used. Previous research showed how the IMC controller can be extended with an anti-windup scheme \cite{SANDIRACONTROLDESIGN} to handle constraints. Also, it was assumed that the orbit response matrix has an accurate circulant symmetry. Measurements show that its structure is affected by the misplacements of certain sensors and actuators.  In order to block-diagonalize the system using a Fourier decomposition, a block-circulant approximation of $R$ is necessary~\cite{SYMSYNCHARXIV}. It remains unclear how this approximation affects the robustness properties of the feedback system. The extension of the controller with an anti-windup scheme and the robustness analysis for a block circulant approximation are currently being considered.

\bibliographystyle{IEEEtran}
\small
\bibliography{IEEEabbrv,bibliography}
\end{document}